\documentclass[12pt]{article}
\usepackage[utf8]{inputenc} 
\usepackage[english]{babel} 
\usepackage{lipsum} 
\usepackage{url} 
\usepackage{authblk}
\usepackage{float}
\usepackage{mathtools}
\usepackage{booktabs} 
\usepackage{fancyhdr} 
\usepackage{hyperref} 
\usepackage{amsmath} 
\usepackage[toc,page]{appendix}
\usepackage{bbold}
\usepackage{blindtext}
\usepackage{caption}
\usepackage{subcaption}
\usepackage{amsthm} 
\usepackage{mathtools} 
\usepackage{amsthm} 
\usepackage{bm} 
\usepackage{braket} 
\usepackage{tikz} 
\usepackage{amssymb} 
\usepackage{amsfonts} 
\newcommand{\R}{\mathbb{R}}

\newcommand{\diff}{\mathrm{d}}

\newcommand{\dt}{\diff t}

\newcommand{\ds}{\diff s}

\DeclareMathOperator{\Tr}{Tr}

\let\oldref\ref
\renewcommand{\ref}[1]{(\oldref{#1})}
\title{\textbf{
\vskip-3cm
Geometric flow equations for the number of space-time dimensions}}
\author[a]{Davide De Biasio}
\author[a]{Julian Freigang}
\author[a,b]{Dieter L\"ust}
\affil[a]{Max--Planck--Institut f\"ur Physik, Werner--Heisenberg--Institut, \newline
F\"ohringer Ring 6, 80805 M\"unchen, Germany}
\affil[b]{Arnold Sommerfeld Center for Theoretical Physics, \newline
Ludwig Maximilians Universit\"at M\"unchen, \newline Theresienstrasse 37, 80333 M\"unchen, Germany}

\usepackage{cite}

\begin{document}

\fancypagestyle{plain}{%
	\fancyhead[R]{LMU-ASC 08/21 \\
MPP-2021-59}
	\renewcommand{\headrulewidth}{0pt}
}
\maketitle

\begin{center}Abstract:
\end{center}
In this paper we consider  new geometric flow equations, called D-flow, which describe the
variation of space-time geometries under the change of the number of dimensions. 
The D-flow is originating from the non-trivial dependence of the volume of space-time manifolds on the number of space-time dimensions and it is driven
by certain curvature invariants. We will work out specific examples of D-flow equations and their solutions for the case of D-dimensional spheres and Freund-Rubin compactified space-time manifolds. 
The discussion of the paper is motivated from recent swampland considerations, where the number $D$ of space-time dimensions is treated as a new swampland parameter.

\newpage

\tableofcontents

\section{Introduction}
The \textit{Swampland Program} \cite{Vafa:2005ui,Palti:2019pca,vanBeest:2021lhn,Brennan:2017rbf} is focused on formulating precise and formal \textit{criteria of demarcation} between those low-energy effective quantum field theories that can be consistently coupled to gravity, admitting an ultraviolet embedding into \textit{quantum gravity}, and those which prevent us from doing so. Given a specific quantum gravity theory, for instance \textit{superstring theory}, the corresponding effective
 low-energy field theories  usually get classified according to the values acquired by some \textit{parameters}. Namely, they can be regarded as points in a generically highly involved and non-trivial, \textit{parameter space}. From this perspective, the portion of the moduli space, for which the effective theories can be consistently coupled to quantum gravity, is referred to as the \textit{Landscape}. In contrast, the \textit{Swampland} is defined as that subset of the theory space populated by models which are structurally precluded from achieving such a quantum gravity embedding. Among the above-mentioned proposed criteria of demarcation, which are usually stated in the form \textit{Swampland Conjectures}, the so-called \textit{Distance Conjecture} \cite{Ooguri:2006in}, relating large distances in parameter space to the appearance of infinite towers of asymptotically massless states, stands out for its strong  geometric connotation and its direct interpretation in the context of superstring Kaluza-Klein compactification.

 Looking at the  distance conjecture from a slightly more general and mathematical point of view, the definition of the distance within the space of background metrics is very closely related to mathematical flow equations in general relativity, like the Ricci flow \cite{hamilton1982,Perelman:2006un,chow2004ricci}, where one follows the flow of a family of 
 metrics with respect to a certain path in field space. In fact,  in a recent series of works \cite{Kehagias:2019akr,Bykov:2020llx,DeBiasio:2020xkv,Luben:2020wix}, we have worked out a close correspondence between the generalized distance conjecture and the geometric flow equations, showing that the fixed points of the Ricci flow with vanishing curvature are typically at infinite distance in the space of geometries that are characterized by geometric parameters like the cosmological constant. Using this observation we conjectured that in general fixed points of the Ricci flow, which are at infinite distance in the background space, are accompanied by an infinite tower of states in quantum gravity. This conjecture was extended for more general gradient flows, like for the combined dilaton-metric flow with a generalized the entropy functional, which then provides a good definition for the distance in the combined field space.

On a different line of thinking it was proposed in \cite{emparan2013large}
and more recently in \cite{Bonnefoy:2020uef} to include the number $D$  of space dimensions into the parameter space
of quantum gravity. This amounts to treat $D$, in addition to the geometric parameters like the radius of a compact space, as new swampland parameter and to include the dependence on $D$
into the distance functionals, which measure the distances between different backgrounds in quantum gravity. This basically amounts to 
examine distances between geometries in gravity with respect to the number of space-time dimensions.

In this paper we try to combine the geometric flow equations with the idea of treating $D$ as independent parameter in quantum gravity.
Namely, we will 
consider  new geometric flow equations, which describe the
variation of space-time geometries under the change of the number of dimensions.  To the best of our knowledge, this flow with respect to $D$, denoted by \textit{D-flow},  
was not discussed before and is introducing a novel mathematical territory in the field of geometric flow equations in gravity. It amounts to 
compute the dependence of the metric on the number of space dimensions and to determine and solve the relevant flow equations,
which describe the D-flow of metric as a function of the relevant curvature invariants. Schematically, the corresponding differential D-flow equation has the following structural form
\begin{equation}
{\rm D-flow}:\qquad{\partial f(g_{\mu\nu})\over \partial D\left(\lambda\right)}=g(R)\, ,
\end{equation}
where $\lambda$ is a flow parameter analogous to the one usually introduced when studying Ricci flow.
Here $f$ is a certain function of the space-time metric and $g$ is a function of certain combinations of the curvature invariants of the geometry. 
As we will discuss, the function $f$ will be closely related to the $D$-dependent
volume of the space-time manifold, whereas in the simplest cases $g$ will be determined by linear and
 quadratic curvature invariants. The appearance of quadratic curvature invariants 
in the flow equation for $D$ is reminiscent to the two-loop graviton $\beta$-function in string theory.
Finally, note that in the D-flow equations we are treating 
  $D$ as a continuous parameter, as it is done e.g. also in the dimensional regularization scheme or also considering the socalled Haussdorf dimensions of certain metric spaces.

\newpage

\section{Derivation of the flow}
Following the standard procedure of deriving the low-energy \textit{graviton} equations of motion emerging from superstring theory by computing the associated $\beta$-function up to a certain power in $\alpha'$ and imposing it to be zero in order to restore the conformal symmetry of the original theory, which was widely explored, for example, in \cite{Callan:1985ia,Callan:1986jb,Foakes:1987gg,Fradkin:1985ys,Graham:1987ep,Grisaru:1986vi,Gross:1986iv,Jack:1989vp}, we turn the \textit{two-loop} expression
\begin{equation}
    \beta_{\mu\nu}=\alpha'R_{\mu\nu}+\frac{\left(\alpha'\right)^2}{2}R_{\mu\alpha\sigma\gamma}R_{\nu}^{\ \alpha\sigma\gamma}
\end{equation}
into the \textit{geometric flow} equation
\begin{equation}\label{Ricciflow}
    \frac{\partial g_{\mu\nu}}{\partial\lambda}\equiv -2\beta_{\mu\nu}=-2\alpha'R_{\mu\nu}-\left(\alpha'\right)^2R_{\mu\alpha\sigma\gamma}R_{\nu}^{\ \alpha\sigma\gamma}\ ,
\end{equation}
where the $\alpha'$-dependence is kept explicit, differently from what is usually done with Ricci flow. This is due to the fact that, here, it can't be removed by a simple \textit{rescaling} of the flow parameter $\lambda$. This choice carries with it two main advantages. First of all, it allows us to quantitatively study how rapidly the flow gets switched-off as the string contributions get smaller, for example by expressing the flow equations it in terms of the ratio between $\sqrt{\alpha'}$ and a typical manifold length scale. Furthermore, it provides us with a powerful tool to properly address the flow of \textit{Ricci-flat} space-time metrics, for which the two-loop term becomes the leading one and source a non-trivial evolution in $\lambda$.
Clearly, there is no known nor direct way to turn the Ricci flow 
equation \eqref{Ricciflow} to a flow equation for the dimension $D$ of the manifold. Therefore, keeping our final purpose in mind, we want to recast it in a suitable form and, then, proceed with \textit{promoting} $D$ to a $\lambda$-dependent quantity. Our first step, starting from \eqref{Ricciflow}, is to observe that it implies the following flow behaviour for the square root of the metric determinant
\begin{equation}
    \frac{\partial\sqrt{g}}{\partial\lambda}=\frac{1}{2\sqrt{g}}\frac{\partial g}{\partial\lambda}\ ,
\end{equation}
where we have chosen to work with Euclidean signature for the equations to be properly defined. For a profound discussion of the many mathematical subtleties that underlie the analysis of geometric flow equations, specifically concerning their \textit{hyperbolicity}, the reader is strongly encouraged to look at the standard references \cite{Topping2006LecturesOT,article,chow2004ricci}.
Using the well-known \textit{Jacobi's} formula
\begin{equation}
    \frac{d}{dt} \det A(t) = \mathrm{tr} \left (\mathrm{adj}(A(t)) \, \frac{dA(t)}{dt}\right )\ ,
\end{equation}
for the derivative of the determinant of a matrix, we get:
\begin{equation}\label{gg}
        \frac{\partial\sqrt{g}}{\partial\lambda}=\frac{1}{2\sqrt{g}}\Tr{\left(gg^{\alpha\mu}\frac{\partial g_{\mu\nu}}{\partial\lambda}\right)}=\frac{\sqrt{g}}{2}\Tr{\left(g^{\alpha\mu}\frac{\partial g_{\mu\nu}}{\partial\lambda}\right)}\ .
\end{equation}
Now, by plugging the flow equation \eqref{Ricciflow} into \eqref{gg}, we are left with
\begin{equation}\label{Y}
    \frac{\partial\sqrt{g}}{\partial\lambda}=-\alpha'\sqrt{g}R-\frac{\left(\alpha'\right)^{2}}{2}\sqrt{g}K\ ,
\end{equation}
with $R$ being the \textit{Ricci scalar} and $K$ the \textit{Kretschmann scalar} associated to $g_{\mu\nu}$. Combining all the metric components into a single equation is surely a step ahead towards our goal, but it is still not enough: so as to achieve a good definition of a flow equation for $D$ and promote it to a continuous modulus of the theory, it is clear that any explicit dependence on the space-time coordinates, whose number is precisely $D$, must be \textit{factored out}. This can be straightforwardly done by integrating both sides of \eqref{Y} on the space-time manifold, over which $g_{\mu\nu}\left(\lambda\right)$ is defined as a family of Riemannian metrics, obtaining
\begin{equation}\label{inter}
    \int\frac{\partial\sqrt{g}}{\partial\lambda}=-\alpha'\int\sqrt{g}R-\frac{\left(\alpha'\right)^{2}}{2}\int\sqrt{g}K\ .
\end{equation}
The integrals, when dealing with a \textit{non-compact} space-time manifold, have to be intended as properly regularised. For example, we might be required to introduce appropriate cut-offs on some coordinates in order to make both sides of the equation finite. At this point, we want to manipulate the left-hand side of \eqref{inter} and express it in a more convenient form. In order to do so, we must first stress that, despite the fact that we will then try to generalise our discussion to a setting in which the manifold's dimension itself can change along the flow, we are still working with the standard form \eqref{Ricciflow} at a fixed value of $D$. Hence, taking the $\lambda$-derivative out of the integral only accounts to the appearance of a \textit{boundary term}, which can be safely dropped. Therefore, we are left with
\begin{equation}\label{vol}
    \frac{\partial\mathcal{V}}{\partial\lambda}=-\alpha'\int\sqrt{g}R-\frac{\left(\alpha'\right)^{2}}{2}\int\sqrt{g}K\ ,
\end{equation}
where the volume $\mathcal{V}$ is defined as:
\begin{equation}
    \mathcal{V}\equiv\int\sqrt{g}\ .
\end{equation}
At this point, we observe that
\begin{equation}
    \mathcal{S}_{1}\equiv\int\sqrt{g}R
\end{equation}
is nothing more than the standard \textit{Einstein-Hilbert} action in Euclidean signature, while 
\begin{equation}
        \mathcal{S}_{2}\equiv\frac{1}{2}\int\sqrt{g}K
\end{equation}
is a higher-order correction of the former. Now, we can see that the final form
\begin{equation}\label{flow}
    \frac{\partial\mathcal{V}}{\partial\lambda}=-\alpha'\mathcal{S}_{1}-\left(\alpha'\right)^{2}\mathcal{S}_{2}\ 
\end{equation}
of the flow equation can be meaningfully generalised to a scenario in which the dimension $D$ of the manifold is not forced to be fixed along the flow, since neither explicit components of the metric nor coordinate-dependent quantities appear in it. This is precisely the route that we will follow. Inspired by \eqref{flow}, we \textit{postulate} the flow equation for $D(\lambda)$ to be:
\begin{equation}\label{Dflow}
    \frac{\diff D}{\diff\lambda}\equiv-\left(\frac{\partial\mathcal{V}}{\partial D}\right)^{-1}\left[\alpha'\mathcal{S}_{1}+\left(\alpha'\right)^{2}\mathcal{S}_{2}\right]\ .
\end{equation}
It must be highlighted that \eqref{flow} and \eqref{Dflow} are \textit{not precisely equivalent}. Namely, if we want to reduce \eqref{Dflow} back to \eqref{flow} we must assume the $D$-derivative of $\mathcal{V}$ to be \textit{finite}, in order to bring it to left-hand side and apply the \textit{chain rule}. As we will see, this is not always the case. Therefore, \eqref{flow} must be regarded as a special case of \eqref{Dflow}. Furthermore, it is clear that the $D$-dependence of $\mathcal{V}$ will descend from our choice of the explicit form of the metric at any given value of $D$, which is \textit{not fixed} by the flow itself. That, in a sense, will be the \textit{input information} allowing us to fix the flow behaviour of the dimension. Usually, such a choice is extremely \textit{natural}, as for the example in which the manifold is taken to always be a $D$-sphere along the flow. Hence, once we have chosen a specific family of metric tensors at different values of $D$, computed the associated scalars $\mathcal{S}_{1}(D)$ and $\mathcal{S}_{2}(D)$ and turned the number of dimensions into a continuous parameter, \eqref{Dflow} must be interpreted as to tool allowing us to find the correct $D\left(\lambda\right)$ so that the $D$-behaviour assumed for $\mathcal{V}$ can be reconciled with \eqref{flow}.

\subsection{Maximally symmetric spaces}
A $D$-dimensional Riemannian manifold is said to be \textit{maximally symmetric} if it possesses the highest allowed number of Killing vectors, namely $D(D+1)/2$. This straightforwardly implies that the associated Riemann curvature tensor gets reduced to the simple form:
\begin{equation}
    R_{\mu\nu\alpha\beta}=\frac{R}{D(D-1)}\left(g_{\mu\alpha}g_{\nu\beta}-g_{\mu\beta}g_{\nu\alpha}\right)\ .
\end{equation}
Therefore, it can be shown that:
\begin{equation}
    K=\frac{2R^{2}}{D(D-1)}\ .
\end{equation}
Because of that, the \textit{volume flow} equation in Euclidean signature gets to be
\begin{equation}
        \frac{\partial\mathcal{V}}{\partial\lambda}=-\int\sqrt{g}\left(\mathcal{R}+\frac{\mathcal{R}^{2}}{D(D-1)}\right)\ ,
\end{equation}
where $\mathcal{R}\equiv\alpha' R$.

\section{Sphere}
Starting from the standard expression for the metric on a $D$-sphere with radius $\rho$, which is indeed a maximally symmetric space, we can straightforwardly derive the expression
\begin{equation}
    R\left(D\right)=\frac{D\left(D-1\right)}{\rho^2}
\end{equation}
for the Ricci scalar and the formula
\begin{equation}
    \mathcal{V}\left(D\right)=\frac{2\pi^{\frac{D+1}{2}}\rho^D}{\Gamma\left(\frac{D+1}{2}\right)}\ ,
\end{equation}
for the $D$-dependence of the volume.
\begin{figure}[ht]
    \centering
    \includegraphics[width=0.8\linewidth]{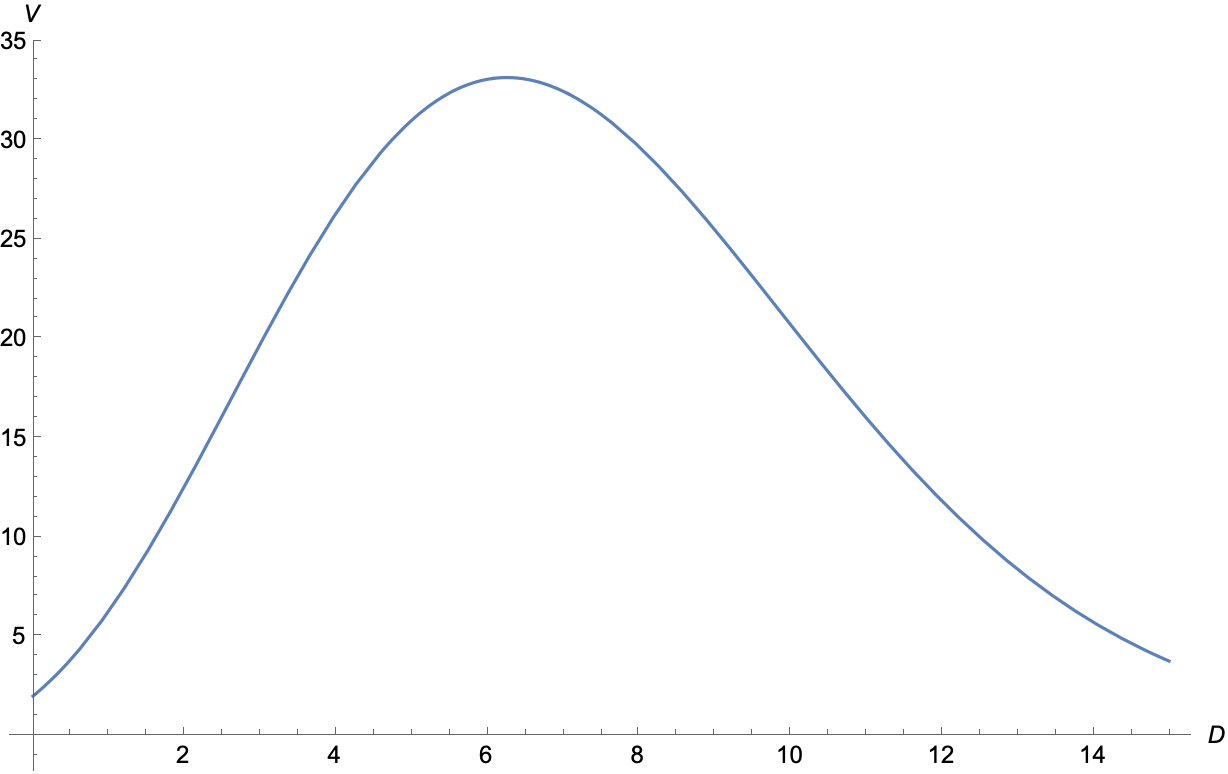}
    \caption{Graph of the volume $\mathcal{V}$ of a $D$-sphere as the number of dimensions $D$ is promoted to a continuous variable.}
\end{figure}\\
At this point, we can proceed towards enforcing the flow equation \eqref{Dflow}. As was widely discussed in the previous section, an explicit solution for $D\left(\lambda\right)$ can only be achieved by specifying from the start a $D$-dependent family of metric tensors. In this example, for any \textit{natural} value of $D$, $g_{\mu\nu}\left(D\right)$ will be nothing more than the metric for a $D$-sphere with radius $\rho$. It must be stressed that we do not need, nor it would have made sense, to specify the form of the metric for any \textit{real} value of $D$. As a matter of fact, it is more than sufficient to turn $D$ into a continuous function $D\left(\lambda\right)$ of the flow parameter after having computed $\mathcal{V}\left(D\right)$, $R\left(D\right)$ and $K\left(D\right)$ as functions of $D$, since all our efforts in rephrasing the flow equation differently were precisely aimed towards allowing us to make $D$ continuous in a consistent way. Therefore, we can now derive the following form for \eqref{Dflow}
\begin{equation}
    \frac{\diff D}{\diff\lambda}=-\frac{D(D-1)}{\mu^{2}}\left(1+\frac{1}{\mu^{2}}\right)\left(\frac{\partial\log{\mathcal{V}}}{\partial D}\right)^{-1}\ ,
\end{equation}
where
\begin{equation}
    \mu\equiv\frac{\rho}{\sqrt{\alpha'}}
\end{equation}
is defined as the ratio between the sphere radius and the length of a string. It is important to highlight the presence of a \textit{deformation} factor
\begin{equation}
    Z\left(\mu\right)\equiv\frac{1}{\mu^{2}}\left(1+\frac{1}{\mu^{2}}\right)
\end{equation}
on the right-hand side of the equation, accounting for the rescaling of the flow produced by the \textit{significance} of string effects with respect to the size of the sphere. Therefore, plotting $Z\left(\mu\right)$ allows us to observe how rapidly the $\lambda$-evolution gets switched off as string effects get negligible, namely as $\sqrt{\alpha'}$ gets much smaller than $\rho$. In particular, we have:
\begin{figure}[ht]
    \centering
    \includegraphics[width=0.8\linewidth]{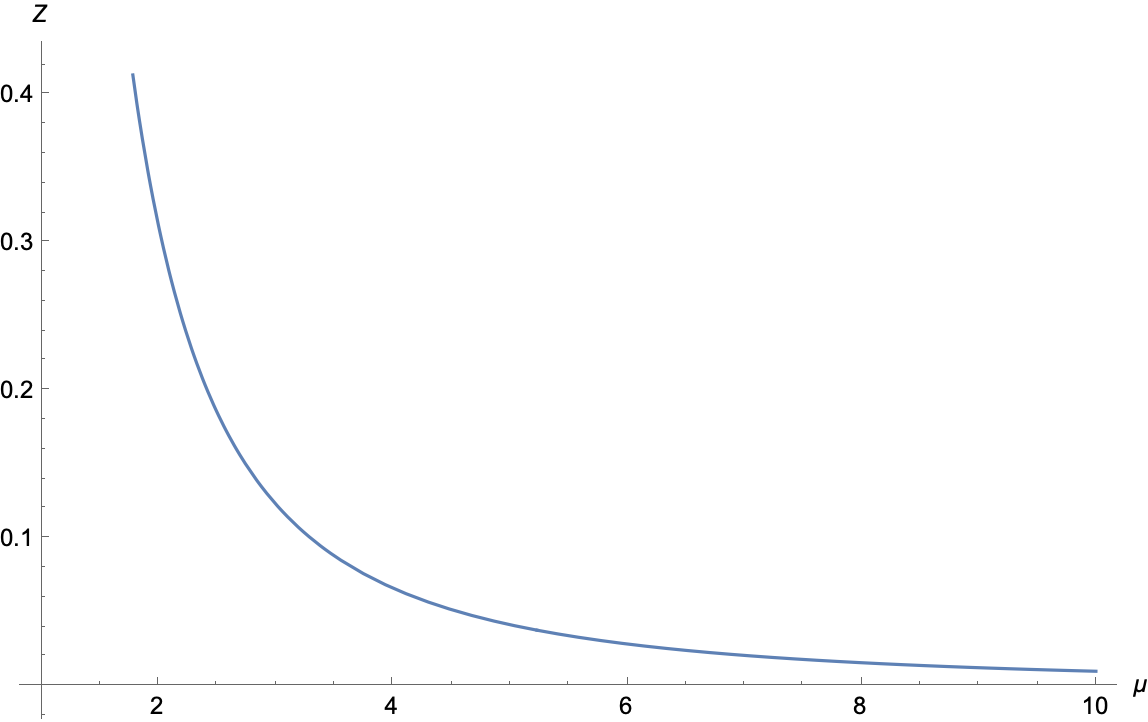}
    \caption{Graph of the deformation factor $Z$ as the scales ratio $\mu$ grows, namely as the sphere radius gets much bigger than the length of a string.}
\end{figure}\\
Given the above considerations, we can now \textit{absorb} $Z(\mu)$ into the flow parameter $\lambda$, defining a new, rescaled, flow parameter
\begin{equation}
    \xi\equiv Z(\mu)\lambda
\end{equation}
and obtaining the following flow equation
\begin{equation}
\begin{split}     
\frac{\diff D}{\diff\xi}=&-D(D-1)\left(\frac{\partial\log{\mathcal{V}}}{\partial D}\right)^{-1}=\\
=&2D(1-D)\left[\log{\left(e\rho^{2}\pi\right)}-\psi_{0}\left(\frac{D+1}{2}\right)\right]^{-1}\ ,
\end{split}
\end{equation}
where $\psi_{0}$ is the \textit{Polygamma} function. Clearly, when writing down the explicit form of the volume $\mathcal{V}$, a dependence on $\rho$ reappears independently from $\sqrt{\alpha'}$. Nevertheless, it is still remarkable that the equation, when $\mathcal{V}$ is implicit and taken as a variable by itself, only depends on the ratio $\mu$.
\subsection{Large-$D$ behaviour}
In this section, we want to study the flow equation at a large number of dimensions, in order to build a better intuition of its asymptotic behaviour. In particular, the explicit form of the volume can be approximated by:
\begin{equation}
    \mathcal{V}\left(D\right)\approx\sqrt{2}e\left(\rho\sqrt{\frac{2\pi e}{D}}\right)^{D}\ .
\end{equation}
Therefore, the flow for $D(\xi)$ can be obtained from
\begin{equation}
    \frac{\diff D}{\diff\xi}=2D(1-D)\left(\log{\frac{2e\pi\rho^2}{D}}\right)^{-1}\approx-2D^2\left(\log{\frac{\sigma}{D}}\right)^{-1}\ ,
\end{equation}
where $\sigma\equiv 2e\pi\rho^2>0$. By rescaling the flow parameter as $\tau\equiv 2\xi\sigma$ and defining the quantity
\begin{equation}
    X\equiv\frac{\sigma}{D}\ ,
\end{equation}
the above equation gets simplified, for very small values of $X$, as:
\begin{equation}
    \frac{\diff X}{\diff\tau}=\frac{1}{\log{X}}\ .
\end{equation}
Therefore, we can simply integrate it
\begin{equation}
    \tau-\tau_{0}=X\left(\log{X}-1\right)-X_{0}\left(\log{X_{0
    }}-1\right)\approx X\log{X}-X_{0}\log{X_{0}}
\end{equation}
and solve for $X$. Starting from a small value of $X_{0}$, corresponding to a large $D$, it is unavoidable to flow towards $X=0$. Indeed, this means that the flow equation forces us to flow towards $D=\infty$.

\subsection{Fixed points}
In the following section, we will look for \textit{fixed points} of the flow. Namely, we will solve the equation
    \begin{equation}\label{fixed}
\frac{\diff D}{\diff\xi}=D(1-D)\left(\frac{\diff\log{\mathcal{V}}}{\diff D}\right)^{-1}=0
\end{equation}
in order to find values of $D$ for which the right-hand side of the flow equation is zero. Then, we will study the stability of such points in detail. Indeed, by plugging-in the explicit form of $\mathcal{V}$, it can be shown that \eqref{fixed} admits two, distinct solutions: $D_{a}=0$ and $D_{b}=1$. Concerning the \textit{stability} of the fixed point $D_{a}$, we observe the following \textit{local} behaviour:\\
\begin{figure}[H]
    \centering
    \includegraphics[width=0.8\linewidth]{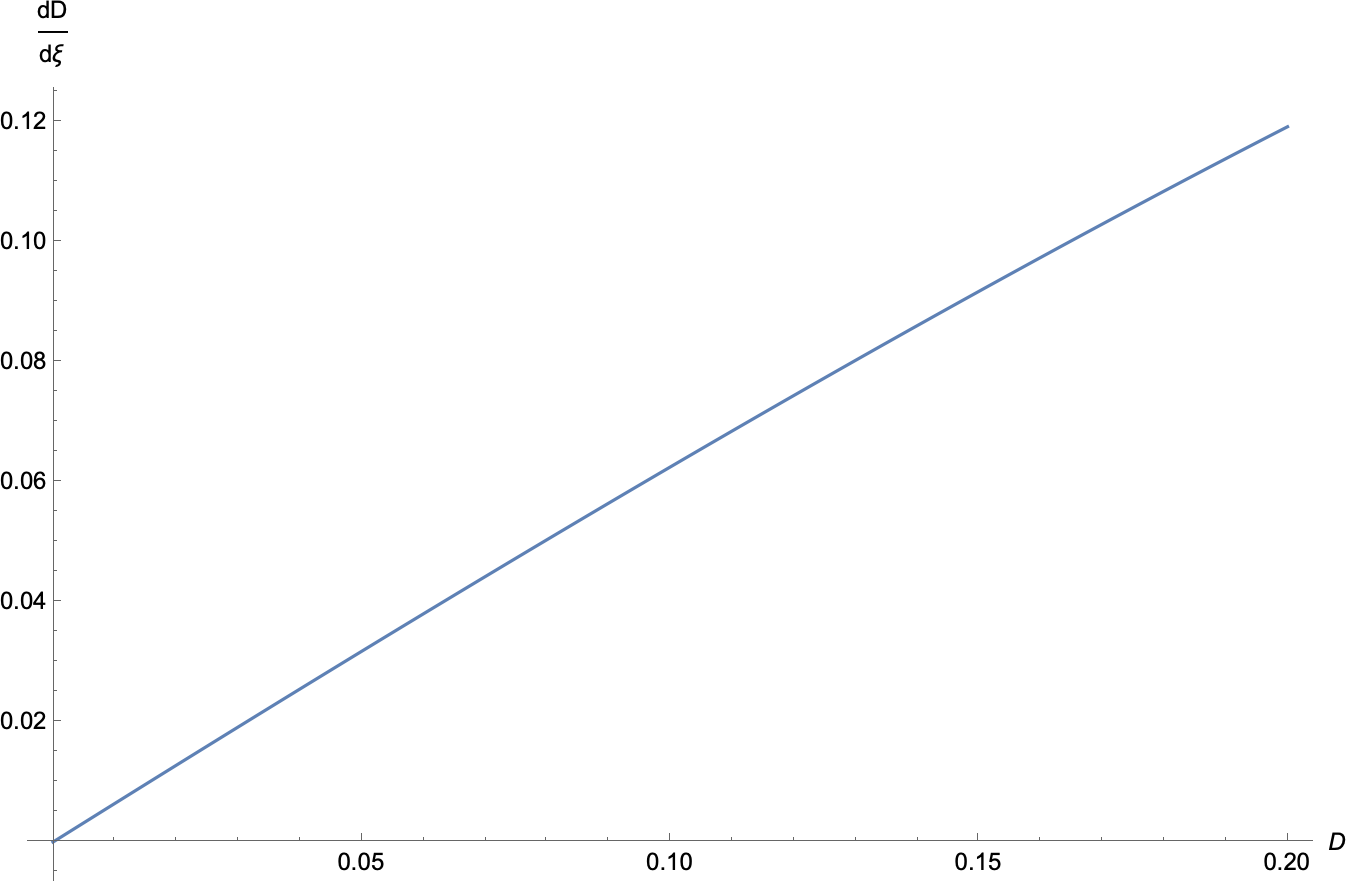}
    \caption{Derivative of $D$ over $\xi$ around $D_{a}$, with $\rho=1$.}
\end{figure}
Therefore, $D_{a}$ is an \textit{unstable} fixed point of the flow. This is due to the fact that $\diff D/\diff\lambda$ is positive for $D$ slightly bigger than $0$: any perturbation is \textit{magnified} by the flow, that brings us far away from $D_{a}$.\\
Concerning, on the other hand, the \textit{stability} of the fixed point $D_{b}$, we observe the following \textit{local} behaviour:\\
\begin{figure}[H]
    \centering
    \includegraphics[width=0.8\linewidth]{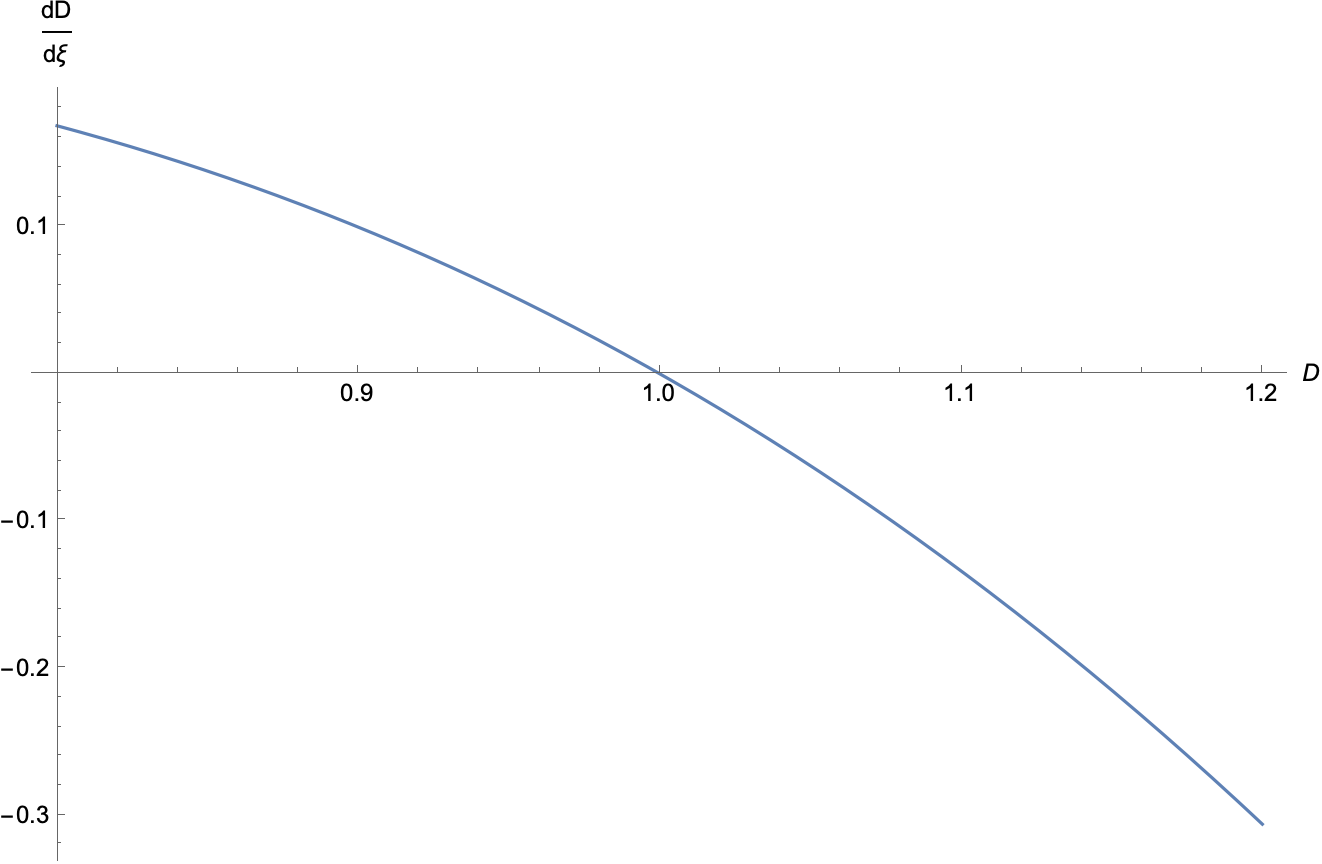}
    \caption{Derivative of $D$ over $\xi$ around $D_{b}$, with $\rho=1$.}
\end{figure}
Hence, $D_{b}$ is a \textit{stable} fixed point of the flow. This is due to the fact that $\diff D/\diff\lambda$ is negative for $D$ slightly bigger than $1$ and positive for $D$ slightly smaller than $1$: any perturbation is \textit{compensated} by the flow, that brings us back to $D_{b}$.
Focusing on $D\in (0,1+\varepsilon)$, with $0<\varepsilon\ll 1$, turning back to the \textit{equivalent} flow equation
    \begin{equation}
\frac{\diff\mathcal{V}}{\diff\lambda}=\frac{\diff D}{\diff\lambda}\frac{\diff\mathcal{V}}{\diff D}=D(1-D)\mathcal{V}
\end{equation}
for the volume, where the chain rule can only be applied since we are in a region where $\mathcal{V}$ is \textit{monotonic} in $D$, and asking ourselves whether the stability of our points is affected by our change of perspective, it is enough to study the sign of $\diff\mathcal{V}/\diff D$. We can straightforwardly observe that we are working in an interval where $\diff\mathcal{V}/\diff D>0$. Thus, this \textit{confirms} the fact that $D_{b}$ is a \textit{stable} fixed point for the volume flow, while $D_{a}$ is \textit{unstable}.

\subsection{Singularities}
\begin{figure}[H]
    \centering
    \includegraphics[width=0.8\linewidth]{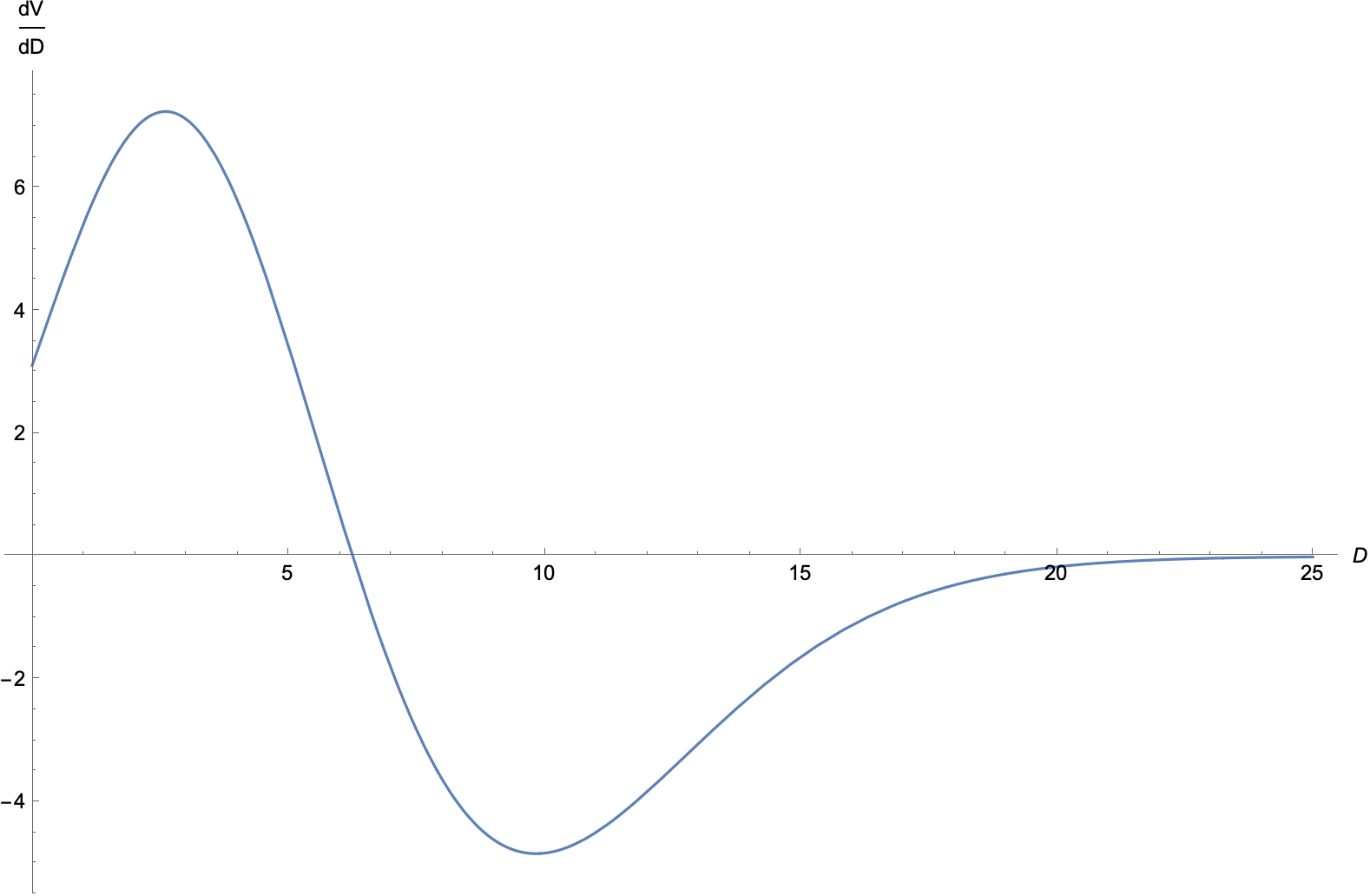}
    \caption{Derivative of $\mathcal{V}$ over $D$, at $\rho=1$ and different values of $D$.}
    \label{Graph}
\end{figure}

At this point, our aim is to locate and study \textit{singular} points along the flow. Namely, we and to find values $\Bar{D}_{a}$ of the dimension for which
\begin{equation}
    \frac{\diff\mathcal{V}}{\diff D}=0\ ,
\end{equation}
namely for which the derivative of $D$ over $\xi$ blows up to infinity. Indeed, it can be easily observed, from the plot presented in Figure \ref{Graph}, that there are two values of $D$ for which the flow gets singular. That is, two \textit{extremal points} of $\mathcal{V}$, when intended explicitly as a function of $D$.
The first one, which we choose to name $\Bar{D}_{a}$, sits at a finite value of $D$, which can be numerically shown to be approximately equal to $6.26$. The second one, oppositely, corresponds to the $D\rightarrow\infty$ limit, which can be consistently referred to as $\Bar{D}_{b}$. While the presence of the former is manifest, the fact that the latter actually corresponds to a singularity as well might still be obscure. In order to dispel any doubts, the limit can be taken by exploiting the large-$D$ approximation of $\mathcal{V}$, producing:
\begin{equation}
    \lim_{D\to\infty}\frac{\diff\mathcal{V}}{\diff D}\approx \lim_{D\to\infty}\left(\frac{2e\pi\rho^{2}}{D}\right)^{\frac{D}{2}}\log{2}=0\ .
\end{equation}
Therefore, it is now clear that the flow equation for $D\left(\xi\right)$ presents two \textit{singular points}. The one at infinity is almost harmless. The one at $\Bar{D}_{a}$, however, is definitely less trivial and requires further attention. In particular, the flow can not be \textit{extended} along the whole real line $\R$ where $D$ is allowed to take values. When the initial point $D_{0}$ is taken to belong to the $(0,\Bar{D}_{a})$ interval, $D\left(\xi\right)$ is confined there too. In an analogous way, choosing for $D_{0}$ a point in $(\Bar{D}_{a},\infty)$ imposes $D\left(\xi\right)$ not to decrease below $\Bar{D}_{a}$.
\begin{figure}[H]
    \centering
    \includegraphics[width=0.8\linewidth]{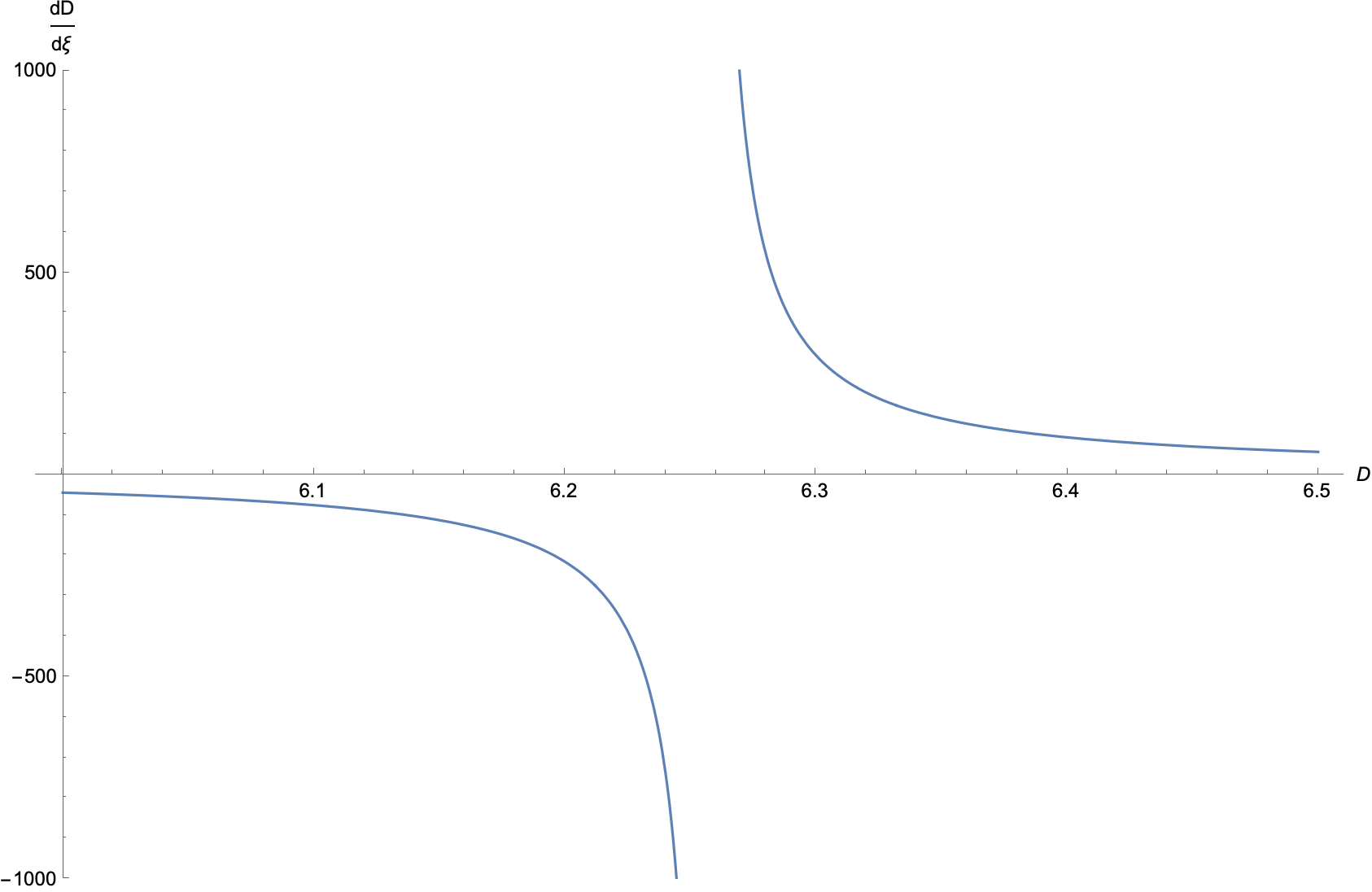}
    \caption{Derivative of $D$ over $\xi$, at $\rho=1$ and around $\Bar{D}_{a}$.}
    \label{Graph2}
\end{figure}
The \textit{stability} of such singularity under small perturbations can be studied by analysing the sign of the $\xi$-derivative of $D$ in its neighbourhood. If $D_{0}$ is chosen to slightly smaller than $\Bar{D}_{a}$, the flow brings $D\left(\xi\right)$ back towards $D_{b}=1$. If, diversely, $D_{0}$ is bigger than $\Bar{D}_{a}$, $D\left(\xi\right)$ flows towards the singular point $\Bar{D}_{b}$ at infinity. Hence, $\Bar{D}_{a}$ is \textit{repulsive}, while $\Bar{D}_{b}$ is attractive.
\section{Anti de Sitter space-time}
In the following section, we compute D-Flow for the case of Anti de Sitter space-time. Namely, we have
\begin{equation}
    \ds^2=-\left(1+\frac{r^2}{\alpha^2}\right)\dt^2+\left(1+\frac{r^2}{\alpha^2}\right)^{-1}\diff r^2+r^2\diff\Omega^2_{D-2}\ ,
\end{equation}
and:
\begin{equation}
    R=\frac{D(1-D)}{\alpha^2}\ .
\end{equation}
Since we are working with a maximally symmetric space, we get:
\begin{equation}
    K=\frac{2R^{2}}{D(D-1)}=\frac{2D(D-1)}{\alpha^4}\ .
\end{equation}
At this point, we introduce a radial cut-off $\Lambda$ and compute the volume enclosed into a sphere with radius $\Lambda$ and centred at $r=0$. Once more, we remove the time integral and get:
\begin{equation}
        \mathcal{V}(D|\Lambda)=\int_{0}^{\Lambda}\diff rr^{D-2}\int_{S_{D-2}}\diff\Omega_{D-2}=\frac{\Lambda^{D-1}}{D-1}\frac{2\pi^{\frac{D-1}{2}}}{\Gamma\left(\frac{D-1}{2}\right)}\ .
\end{equation}
By taking $\Lambda=\alpha$, we arrive to
\begin{equation}
    \frac{\diff D}{\diff\lambda}=\frac{D(D-1)}{\sigma^{2}}\left(1-\frac{1}{\sigma^{2}}\right)\left(\frac{\partial\log{\mathcal{V}}}{\partial D}\right)^{-1}\ ,
\end{equation}
with:
\begin{equation}
    \sigma\equiv\frac{\alpha}{\sqrt{\alpha^{'}}}\ .
\end{equation}
Hence, the deformation factor $Z(\sigma)$ is slightly different from the one we had for the sphere. In particular, the flow gets weak again when approaching $\sigma\sim 1$. This is specifically due to the sign of $R$. In order to better investigate the shape of $Z$, it can be interesting to include higher terms.
By reabsorbing the $\sigma$-dependent factor into the flow equation, we have a fixed point at $D=0$ and one at $D=1$.
\begin{figure}[H]
    \centering
    \includegraphics[width=0.8\linewidth]{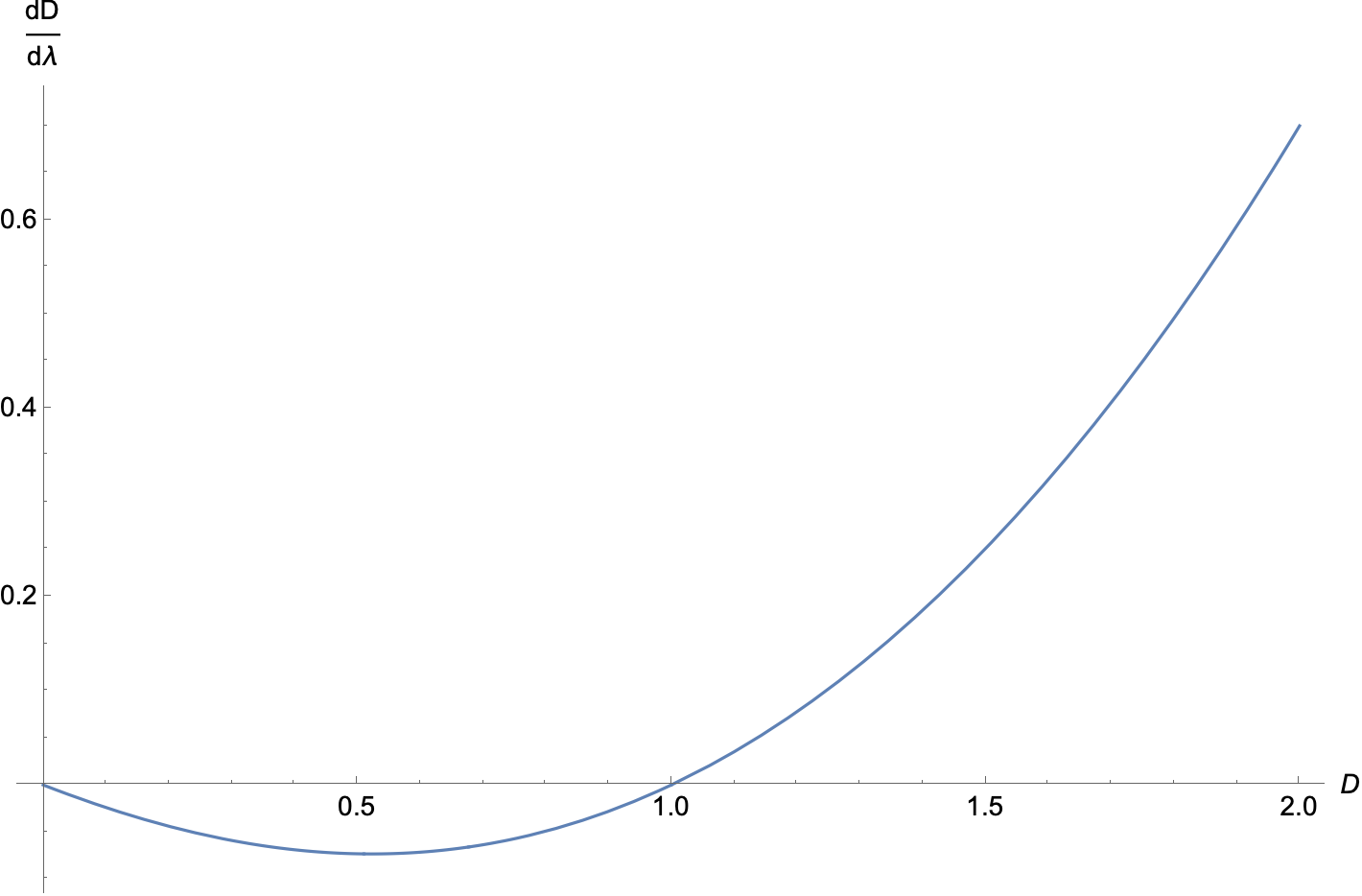}
    \caption{Derivative of $D$ over $\lambda$ versus $D$, at $\alpha=1$.}
    \label{Graph3}
\end{figure}
Therefore, for $D>1$ we are pushed towards $D=\infty$. Since our solution is defined for $D\geq 2$, we are \textit{always} pushed to $\infty$. 
\newpage
\section{Freund-Rubin Compactification}
In the following section, we consider a particular setting for superstring theory compactification on a sphere. Namely, we analyse Freund-Rubin compactification, as presented in \cite{Bonnefoy:2020uef}, and write down its associated D-Flow equation. Since we are dealing with a non-warped product manifold
\begin{equation}
    \mathcal{M}_{D}=AdS_d\times S^{d'}\ ,
\end{equation}
with $D\equiv d+d'$, we will be left with an evolution equation for $D$, which will have to be translated into a flow for $d$ and one for $d'$. In order to do so, our \textit{underdetermined} system forces us to impose a further condition on either $d$ or $d'$. Since the low energy effective field theory description we are interested in is expected to live on AdS space-time, the most natural choice is to assume its dimension to be fixed and move the whole flow dependence to teh compact manifold dimension $d'$. 
\subsection{Description of the Setup}
The \textit{Freund-Rubin} compactification is a non-warped product manifold $\mathcal{M}_D = AdS_d$ x $S^{d'}$ with $D=d+d'$, where $G$ is the metric of $\mathcal{M}_D$, $g_A$ is the metric of AdS space-time and $g_S$ is the metric of the sphere. It can be obtained as a solution to the $D$-dimensional Einstein field equations without cosmological constant in the presence of a $d$-form field strength localised on AdS
\begin{equation}
F_{\mu_1 ... \mu_d} = \frac{\epsilon_{\mu_1...\mu_d}}{\sqrt{-g_A}}f\ ,
\end{equation}
where $f$ is a constant with units of mass squared.\\
The presence of the field strength modifies the Ricci scalars of the AdS- space $\mathcal{R}_A$ and the Ricci scalar of the sphere $\mathcal{R}_S$ to
\begin{eqnarray}
\mathcal{R}_A &=  - \frac{d(d'-1)}{D-2}f^2\ ,\\
\mathcal{R}_S &=  \frac{d'(d-1)}{D-2}f^2\ .
\end{eqnarray}
This also implies a correlation between the AdS radius $R_A$ and the sphere radius $R_S$:
\begin{equation}
(d-1)R_S = (d'-1)R_A\ . \label{relation}
\end{equation}
Furthermore, the coordinates of the two subspaces do not mix. Hence, the volume just factorises as
\begin{equation}
\mathcal{V}_D = \int_{\mathcal{M}_D} \sqrt{-G} = \int_{AdS_d} \sqrt{-g_A} \int_{S^{d'}} \sqrt{g_S} = \mathcal{V}_{AdS_d} \mathcal{V}_{S^{d'}}
\end{equation}
and the total Ricci scalar is just the sum of the Ricci scalars of the subspaces:
\begin{equation}
\mathcal{R}_D = \mathcal{R}_A + \mathcal{R}_S  =\frac{f^2 (d-d')}{D-2}\ . \label{RicciScalar}
\end{equation}
\subsection{Flow }
Constructing the D-Flow equation for this example, we only focus on the first order contribution in $\alpha'$. Hence, such parameter can be reabsorbed into the flow parameter $\lambda$. Since the Ricci scalar does not depend on the coordinates, the D-Flow equation simply reduces to:
\begin{equation}
\frac{d}{d\lambda} D(\lambda) = - \mathcal{R}_D \left( \frac{\partial \log \mathcal{V}_D}{\partial D} \right)^{-1}\ .
\end{equation}
As was previously discussed, we now impose $d$ to be fixed and move the whole flow dependence to $d'$. Hence, we consider $\mathcal{V}$ as a function of $d'$ and obtain the following expression
\begin{equation}
    \frac{d}{d\lambda} d'(\lambda) = - \mathcal{R}_D \left( \frac{\partial \log \mathcal{V}_{D}}{\partial d'} \right)^{-1}\ ,
\end{equation}
where:
\begin{align}
\mathcal{V}_{S^{d'}} &= \frac{R_S^{d'}2 \pi^{\frac{d'+1}{2}}}{\Gamma\left( \frac{d'+1}{2} \right)}\\
\frac{\partial \log \mathcal{V}_D}{\partial d'}&= \frac{\partial \log \mathcal{V}_{S^{d'}}}{\partial d'}= \frac{1}{2}\left(\log( \pi R_S^2)-\psi_0\left( \frac{d'+1}{2}\right)\right),
\end{align}
The resulting flow equation for $d'$ is:
\begin{equation}
\frac{d}{d\lambda} d'(\lambda) = - \frac{2 f^2 (d-d')}{d+d'-2}\frac{1}{\log( \pi R_S^2)-\psi_0\left( \frac{d'+1}{2}\right)}\ . \label{flow2}
\end{equation}
It must be stressed that the above derivation assumes both $R_{S}$ and $R_{A}$ to be fixed along the flow, unavoidably violating the condition expressed in \eqref{relation}. Otherwise, we can choose to impose it and allow (at least) one of the radii to change with $d'$. This option will be discussed later.
First of all, it can be straightforwardly observed that the above expression has two fixed points: one at $d'=d$ and one at $d'=\infty$.
By studying the sign of the RHS of \eqref{flow2}, we can analyse the character of such points. In particular, we observe that:
\begin{itemize}
    \item $d'=d $ is an \textit{unstable} fixed point. By taking $d'$ slightly smaller than $d$, we get pushed to $0$. By taking, on the other hand, $d'$ slightly larger than $d$, we get pushed to $\infty$.
    \item $d'=\infty $ is a \textit{stable} fixed point. By taking $d'>d$, we always get pushed to $\infty$.
\end{itemize}
\subsubsection{Fixed AdS radius}
In the following discussion, we assume the radius $R_{A}$ of AdS space-time to be fixed, impose \eqref{relation} and introduce a $\lambda$-dependence in the sphere radius $R_{S}$. In particular, we have:
\begin{eqnarray}
R_{S}(\lambda)=\frac{d'(\lambda)-1}{d-1}R_{A}\ .
\end{eqnarray}
This unavoidably alters the flow equation for $d'$, which takes the form:
\begin{equation}
   \frac{d}{d\lambda} d'(\lambda)= - \frac{2 f^2 (d-d')}{d+d'-2}\frac{1}{2\frac{d'}{d'-1}+\log{\left(\pi R_{S}(\lambda)^2\right)}-\psi_{0}\left(\frac{d'+1}{2}\right)}\ .
\end{equation}
The expression for $R_S(\lambda)$ is the one presented above.
As can be clearly observed, a new term has been introduced in the denominator. It doesn't alter the unstable behaviour of the fixed point at $d'=d$, nor the stability of the one at $d'=\infty$. Nevertheless, it introduces a further \textit{stable} fixed point at $d'=1$. The sphere radius, in correspondence to the fixed points, assumes three peculiar values:
\begin{itemize}
    \item At $d'=1$, the sphere turns into a $1$-dimensional circle. Hence, the whole computation of the curvature breaks down and the flow gets pathological.
    \item At $d'=d$, $R_S$ is equal to $R_A$.
    \item At $d'=\infty$, $R_S$ grows too towards $\infty$. Hence, KK states are expected to produce a tower of massless states.
\end{itemize}
The picture emerging when both the AdS dimension and radius are kept fixed, while varying the internal sphere dimension and size, can be summarised as follows. We have an unstable fixed point at $d=d'$, with $R_A=R_S$, where the theory seems to be consistent. As soon as a small perturbation of the sphere dimension is introduced, we get either pushed towards $d'=1$, where our flow equations get pathological, or towards $d'=\infty$, where an infinite tower of states is expected to appear in the spectrum. Concerning the specific scaling behaviour of KK modes, it was presented in \cite{Bonnefoy:2020uef} as
\begin{equation}
    m^2_{KK}(l=1)=-2\frac{(d-1)d'}{(d-2)(d'-1)^2}\Lambda_{d}\ ,
\end{equation}
where $l$ labels KK momentum and, in our derivation, the cosmological constant $\Lambda_{d}$ of the AdS effective field theory was chosen to not to vary along the flow, as the whole $\lambda$-dependence was moved to parameters of the internal dimensions. Therefore, it can be clearly observed that such states get asymptotically massless as we flow towards $d'=\infty$. In particular, following the standard discussions of the Swampland Distance Conjecture (SDC), we expect the flow to be provided with an appropriate notion of distance $\Delta\left(\lambda_{0},\lambda\right)$, so that we asymptotically have:
\begin{equation}
    m^2_{KK}\left(\lambda\right)\sim m^2_{KK}\left(0\right)e^{-\alpha\Delta\left(\lambda_{0},\lambda\right)}\ .
\end{equation}
In our example, focusing on the $d'=\infty$ limit, this would translate into identifying the asymptotic behaviour of the distance with:
\begin{equation}
    \alpha\Delta\left(\lambda_{0},\lambda\right)\sim\log{\frac{m^2_{KK}\left(0\right)}{m^2_{KK}\left(\lambda\right)}}=\log{\frac{d'_0(d'_\lambda-1)^2}{d'_\lambda(d'_0-1)^2}}\sim\log{d'_\lambda}\ .
\end{equation}
This clearly doesn't uniquely fix an appropriate notion of $\Delta$, as it only regards its long-distance behaviour. Nevertheless, it fits the standard expectation that the distance should grow proportionally with the logarithm of the dimension and allows to observe that $d'=\infty$ is at infinite distance from the unstable, consistent fixed point at $d'=d$. 
\section{The Swampland}
As was briefly discussed in the introductory section, some literature \cite{DeBiasio:2020xkv,Kehagias:2019akr,Bykov:2020llx,Luben:2020wix} has been recently produced in the attempt of realising the Swampland Distance Conjecture \cite{Ooguri:2006in} via geometric flow equations. The SDC was originally motivated by observing that the physics of $d$-dimensional effective field theories, derived by compactifying $D=(d+d')$-dimensional superstring theory, strongly depends on the moduli defining the shape of the internal dimensions. In $S^1$ compactification, for instance, the $9$-dimensional effective theory \cite{Palti:2019pca} is characterised by two infinite towers of states, getting asymptotically massless when the radius of the circle is, respectively, sent to $0$ or $\infty$. Therefore, after having turned a string-motivated geometric flow into an evolution equation for the number $D$ of space-time dimensions, we have decided to consider the specific setting of Freund-Rubin Compactification. In the spirit outlined above, we have chosen to consider the AdS effective theory parameters to be fixed with $\lambda$, while the whole flow dependence was moved to the internal generalised moduli: namely, to the radius and the dimension of the internal sphere. This procedure clearly fits into the typical settings realising the SDC, as the simple example of circle compactification. This way we have derived that, when slightly perturbing the fixed point at $d=d'$, one is dragged towards $d'=1$ or $d'=\infty$. At $d'=\infty$, the most interesting of the two, the AdS effective theory becomes inconsistent due to a tower of KK states getting massless. Indeed, this allowed us to derive a rough estimate of the asymptotic behaviour of the natural notion of distance with which our generalised moduli space should be equipped with. This was enough to establish that $d'=\infty$ lies infinitely far along the flow. The precise definition of a distance goes beyond the scope of this work and would likely require a deeper analysis of the geometric structure of such moduli space.

\section{Conclusions}
Starting from the expression for the \textit{graviton} $\beta$-function coming from the superstring theory worldsheet $\sigma$-model, a two-loop refined version of \textit{Ricci flow} equation was derived. Thus, the associated flow equation for the volume $\mathcal{V}$ of the manifold on which the flow takes place was explicitly constructed. Exploiting its suitable form, namely it being independent from the coordinates on the manifold and not consisting in a system of component-by-component equations whose number straightforwardly comes from the dimension of the manifold, $D$ was generalised to a continuous parameter and provided with an analogous flow. It must be stressed that the evolution equation for $D$ only corresponds to the one for $\mathcal{V}$ when it is non-singular: in this sense, the equation for $D$ can be regarded as a generalisation of the framework from which we started. Thereafter, the explicit example of a family of $D$-spheres was studied, highlighting the interesting behaviour around \textit{fixed} and \textit{singular} points. In both cases, the \textit{attractive} or \textit{repulsive} behaviour of the singled-out values of $D$ was studied in detail. 
Thereafter, the specific example of Freund-Rubin compactification was analysed in great detail, finding some remarkable similarities with the behaviour one would expect from the SDC.\\

Let us also briefly compare the results of this paper with the swampland constraints for large $D$ gravity theories, which were discussed in \cite{Bonnefoy:2020uef}.
As we have seen here, the D-flow becomes singular in the limit $D\rightarrow\Bar{D}_{b}=\infty$, where this singular point of the D-flow was shown to be attractive.
On the other hand, following the swampland arguments
of \cite{Bonnefoy:2020uef} for $AdS_D\times S^D$ backgrounds, there is an upper bound on the number of space-time dimensions, namely $D\leq\rho^2$, where $\rho$ is the radius of the D-sphere.
This bound arises, since 
in this limit the  volume of the D-sphere is shrinking and the mass scale of the associated Kaluza-Klein tower becomes super-Planckian. 
However, since the singular point $\Bar{D}_{b}$ possesses very large curvature, additional curvature invariants should be added to the D-flow equations. This might alter some of the conclusions about the stability of this singular point.
\section{Acknowledgements}
We thank Cesar Gomez and Alex Kehagias 
for the useful comments.\\ The work of D.L. is supported  by the Origins Excellence Cluster.
\newpage
\bibliographystyle{utphys}
\bibliography{bibliogra.bib}

\providecommand{\href}[2]{#2}\begingroup\raggedright\begin{thebibliography}{10}

\bibitem{Vafa:2005ui}
C.~Vafa, ``{The String landscape and the swampland},'' 9, 2005.

\bibitem{Palti:2019pca}
E.~Palti, ``{The Swampland: Introduction and Review},''
  \href{http://dx.doi.org/10.1002/prop.201900037}{{\em Fortsch. Phys.}
  {\bfseries 67} no.~6, (2019) 1900037},
  \href{http://arxiv.org/abs/1903.06239}{{\ttfamily arXiv:1903.06239
  [hep-th]}}.

\bibitem{vanBeest:2021lhn}
M.~van Beest, J.~Calder\'on-Infante, D.~Mirfendereski, and I.~Valenzuela,
  ``{Lectures on the Swampland Program in String Compactifications},''
  \href{http://arxiv.org/abs/2102.01111}{{\ttfamily arXiv:2102.01111
  [hep-th]}}.

\bibitem{Brennan:2017rbf}
T.~D. Brennan, F.~Carta, and C.~Vafa, ``{The String Landscape, the Swampland,
  and the Missing Corner},'' \href{http://dx.doi.org/10.22323/1.305.0015}{{\em
  PoS} {\bfseries TASI2017} (2017) 015},
  \href{http://arxiv.org/abs/1711.00864}{{\ttfamily arXiv:1711.00864
  [hep-th]}}.

\bibitem{Ooguri:2006in}
H.~Ooguri and C.~Vafa, ``{On the Geometry of the String Landscape and the
  Swampland},'' \href{http://dx.doi.org/10.1016/j.nuclphysb.2006.10.033}{{\em
  Nucl. Phys. B} {\bfseries 766} (2007) 21--33},
  \href{http://arxiv.org/abs/hep-th/0605264}{{\ttfamily arXiv:hep-th/0605264}}.

\bibitem{hamilton1982}
R.~S. Hamilton, ``Three-manifolds with positive ricci curvature,''
  \href{http://dx.doi.org/10.4310/jdg/1214436922}{{\em J. Differential Geom.}
  {\bfseries 17} no.~2, (1982) 255--306}.
  \url{https://doi.org/10.4310/jdg/1214436922}.

\bibitem{Perelman:2006un}
G.~Perelman, ``{The Entropy formula for the Ricci flow and its geometric
  applications},'' \href{http://arxiv.org/abs/math/0211159}{{\ttfamily
  arXiv:math/0211159}}.

\bibitem{chow2004ricci}
B.~Chow and D.~Knopf, {\em The Ricci Flow: An Introduction: An Introduction},
  vol.~1.
\newblock American Mathematical Soc., 2004.

\bibitem{Kehagias:2019akr}
A.~Kehagias, D.~L{\"u}st, and S.~L{\"u}st, ``{Swampland, Gradient Flow and
  Infinite Distance},'' \href{http://dx.doi.org/10.1007/JHEP04(2020)170}{{\em
  JHEP} {\bfseries 04} (2020) 170},
  \href{http://arxiv.org/abs/1910.00453}{{\ttfamily arXiv:1910.00453
  [hep-th]}}.

\bibitem{Bykov:2020llx}
D.~Bykov and D.~L{\"u}st, ``{Deformed $\sigma$-models, Ricci flow and Toda
  field theories},'' \href{http://arxiv.org/abs/2005.01812}{{\ttfamily
  arXiv:2005.01812 [hep-th]}}.

\bibitem{DeBiasio:2020xkv}
D.~De~Biasio and D.~L{\"u}st, ``{Geometric Flow Equations for Schwarzschild-AdS
  Space-Time and Hawking-Page Phase Transition},''
  \href{http://dx.doi.org/10.1002/prop.202000053}{{\em Fortsch. Phys.}
  {\bfseries 68} no.~8, (2020) 2000053},
  \href{http://arxiv.org/abs/2006.03076}{{\ttfamily arXiv:2006.03076
  [hep-th]}}.

\bibitem{Luben:2020wix}
M.~L{\"u}ben, D.~L{\"u}st, and A.~R. Metidieri, ``{The Black Hole Entropy
  Distance Conjecture and Black Hole Evaporation},''
  \href{http://dx.doi.org/10.1002/prop.202000130}{{\em Fortsch. Phys.}
  {\bfseries 69} no.~3, (2021) 2000130},
  \href{http://arxiv.org/abs/2011.12331}{{\ttfamily arXiv:2011.12331
  [hep-th]}}.

\bibitem{emparan2013large}
R.~Emparan, R.~Suzuki, and K.~Tanabe, ``The large d limit of general
  relativity,'' {\em Journal of High Energy Physics} {\bfseries 2013} no.~6,
  (2013) 9.

\bibitem{Bonnefoy:2020uef}
Q.~Bonnefoy, L.~Ciambelli, D.~L{\"u}st, and S.~L{\"u}st, ``{The Swampland at
  Large Number of Space-Time Dimensions},''
  \href{http://arxiv.org/abs/2011.06610}{{\ttfamily arXiv:2011.06610
  [hep-th]}}.

\bibitem{Callan:1985ia}
C.~G. Callan, Jr., E.~J. Martinec, M.~J. Perry, and D.~Friedan, ``{Strings in
  Background Fields},''
  \href{http://dx.doi.org/10.1016/0550-3213(85)90506-1}{{\em Nucl. Phys. B}
  {\bfseries 262} (1985) 593--609}.

\bibitem{Callan:1986jb}
C.~G. Callan, Jr., I.~R. Klebanov, and M.~J. Perry, ``{String Theory Effective
  Actions},'' \href{http://dx.doi.org/10.1016/0550-3213(86)90107-0}{{\em Nucl.
  Phys. B} {\bfseries 278} (1986) 78--90}.

\bibitem{Foakes:1987gg}
A.~P. Foakes and N.~Mohammedi, ``{An Explicit Three Loop Calculation for the
  Purely Metric Two-dimensional Nonlinear $\sigma$ Model},''
  \href{http://dx.doi.org/10.1016/0550-3213(88)90696-7}{{\em Nucl. Phys. B}
  {\bfseries 306} (1988) 343--361}.

\bibitem{Fradkin:1985ys}
E.~S. Fradkin and A.~A. Tseytlin, ``{Quantum String Theory Effective Action},''
  \href{http://dx.doi.org/10.1016/0550-3213(85)90559-0}{{\em Nucl. Phys. B}
  {\bfseries 261} (1985) 1--27}. [Erratum: Nucl.Phys.B 269, 745--745 (1986)].

\bibitem{Graham:1987ep}
S.~J. Graham, ``{Three Loop Beta Function for the Bosonic Nonlinear $\sigma$
  Model},'' \href{http://dx.doi.org/10.1016/0370-2693(87)91052-5}{{\em Phys.
  Lett. B} {\bfseries 197} (1987) 543--547}.

\bibitem{Grisaru:1986vi}
M.~T. Grisaru and D.~Zanon, ``{$\sigma$ Model Superstring Corrections to the
  Einstein-hilbert Action},''
  \href{http://dx.doi.org/10.1016/0370-2693(86)90765-3}{{\em Phys. Lett. B}
  {\bfseries 177} (1986) 347--351}.

\bibitem{Gross:1986iv}
D.~J. Gross and E.~Witten, ``{Superstring Modifications of Einstein's
  Equations},'' \href{http://dx.doi.org/10.1016/0550-3213(86)90429-3}{{\em
  Nucl. Phys. B} {\bfseries 277} (1986) 1}.

\bibitem{Jack:1989vp}
I.~Jack, D.~R.~T. Jones, and N.~Mohammedi, ``{A Four Loop Calculation of the
  Metric Beta Function for the Bosonic $\sigma$ Model and the String Effective
  Action},'' \href{http://dx.doi.org/10.1016/0550-3213(89)90422-7}{{\em Nucl.
  Phys. B} {\bfseries 322} (1989) 431--470}.

\bibitem{Topping2006LecturesOT}
P.~M. Topping, ``Lectures on the ricci flow,''
\newblock 2006.

\bibitem{article}
C.~Mantegazza, G.~Catino, L.~Cremaschi, Z.~Djadli, and L.~Mazzieri, ``The
  ricci-bourguignon flow,''
  \href{http://dx.doi.org/10.2140/pjm.2017.287.337}{{\em Pacific Journal of
  Mathematics} {\bfseries 287} (07, 2015) }.

\end{thebibliography}\endgroup

\end{document}